\documentclass[preprint,aps]{revtex4}

\usepackage{amsmath}
\usepackage{graphicx}
\usepackage{bm}
\usepackage{fancyhdr}
\usepackage{amsmath}
\usepackage{graphicx}
\usepackage{amsfonts}
\usepackage{amsbsy}
\usepackage{amssymb}
\usepackage[mathscr]{eucal}

\begin{document}
	
\title{The Electron-Phonon Coupling Constant for Single-Layer Graphene on Metal Substrates Determined from He Atom Scattering }

\author{G. Benedek$^{a,b}$, J. R. Manson$^{a,c}$, Salvador Miret-Art{\'e}s$^{a,d}$}
	
\affiliation{$^a$ Donostia International Physics Center (DIPC),~Paseo Manuel de Lardiz{{a}}bal,~4,~20018 Donostia-San Sebastian, Spain}

\affiliation{$^b$ Dipartimento di Scienza dei Materiali,~Universit{\`a} di Milano-Bicocca,~Via Cozzi 55,~20125 Milano, Italy }

\affiliation{$^{c}$ Department of Physics and Astronomy, Clemson University, Clemson, South Carolina 29634, U.S.A.}

\affiliation{$^d$ Instituto de F\'isica Fundamental, Consejo Superior de Investigaciones Cient\'ificas, Serrano 123, 28006 Madrid, Spain}

\begin{abstract}
Recent theory has demonstrated that the value of the electron-phonon coupling strength $\lambda$ can be extracted directly from the 
thermal attenuation (Debye-Waller factor) of Helium atom scattering reflectivity. This theory is here extended to multivalley 
semimetal systems 
and applied to the case of graphene on different metal substrates and graphite. It is shown that $\lambda$ rapidly increases for 
decreasing graphene-substrate binding strength. 
Two different calculational models  are considered which produce qualitatively similar results for the dependence of $\lambda$ on binding strength.
These models predict, respectively, values of $\lambda_{HAS} = 0.89$ and 0.32 for a hypothetical flat free-standing single-layer  graphene with cyclic boundary conditions.
The method is suitable for analysis and characterization of not only the graphene overlayers 
considered here, but also other layered systems such as twisted graphene bilayers.
\end{abstract}

\maketitle

\section{Introduction}

Current interest in single-layer graphene supported on metal substrates has led several experimental groups to investigate a number of such 
systems with He atom scattering (HAS), \cite{Farias-2DMat-18,Kondo-09,Yamada-15,Farias-Carbon-18,Sibener-16,Farias-Carbon-15,Farias-Carbon-15-2,Farias-Carbon-16,
Farias-17,Farias-PRB-11,Farias-17x,Tam-15} as well as the surface of clean highly ordered pyrolytic graphite (HOPG), C(0001).\cite{Farias-2DMat-18,Kondo-09,Yamada-15}
Two of these systems, namely graphene (Gr) on Ni(111) and Gr/Ru(111) have also been investigated with Ne atom scattering. \cite{Farias-PRB-11,Farias-17x}  In all of these systems high quality data are available for the thermal attenuation of the specular diffraction peak 
over a large range of temperatures.
Such thermal attenuation measurements are interesting because it  has been shown that they can be used to extract values of the electron-phonon coupling constant $\lambda$ (also known as the mass correction factor) in the surface region.

The ability of atom scattering to measure $\lambda$ relies on the fact that colliding atoms are repelled by the surface electron density arising from electronic states near the Fermi level, and energy is exchanged with the phonon gas primarily via the electron-phonon coupling.  The electron-phonon coupling constant $\lambda$ at the surface is defined as the average
$\lambda = \langle \lambda_{{\bf Q},\nu} \rangle = \sum_{{\bf Q},\nu} \lambda_{{\bf Q},\nu}/3N$ 
over the phonon mode-dependent contributions $\lambda_{{\bf Q},\nu} $
where ${\bf Q}$ is the phonon parallel wave vector, $\nu$ the branch index, and $N$ is the total number of atoms of the crystal. \cite{Allen}

It has been theoretically demonstrated that the 
intensity of 
peak features due to specific $({\bf Q}, \nu)$ phonon modes as observed in inelastic He atom scattering spectra are individually proportional to their corresponding $\lambda_{{\bf Q},\nu}$. \cite{Skl,Benedek-14}
This prediction has been verified through detailed comparisons of calculations with experimental He scattering measurements of multiple layers of Pb on a Cu(111) substrate.\cite{Skl,Benedek-14}
Since the thermal attenuation of any quantum peak feature in the atom scattering spectra is due to an average over the mean square displacement of all phonon modes
weighted by the respective electron-phonon coupling, 
it is not surprising that such attenuation can be related to
the average
$\lambda_{HAS} = \langle \lambda_{{\bf Q}, \nu} \rangle$. \cite{Manson-JPCL-16,JPCL2,Manson-SurfSciRep,JPCL-2020} This will be discussed in more detail below in Section~\ref{Theory} 
where the theory is briefly outlined.  In Section~\ref{Graphite} values of $\lambda_{HAS}$ are obtained from the available He atom scattering data on C(0001). Section~\ref{Graphene} presents
an analysis of $\lambda_{HAS}$ from the available data 
on Gr adsorbed on close-packed metal substrates by analyzing the thermal attenuation of the specular He atom scattering peaks. 
Two different calculational models are considered and both show that the 
$\lambda_{HAS}$ values exhibit a similar and interesting relationship when compared with the relative binding strengths of the graphene to the metal surface. For binding strengths approaching zero this leads to predictions for the value of $\lambda_{HAS}$  for free-standing single-layer graphene.
Finally, a discussion is presented of the shear-vertical (ZA) mode of substrate-supported thin layers such as Gr/metals, and how it compares to the flexural mode of a thin flake of the same unsupported two dimensional (2D) material with free-boundary conditions. A summary and a few conclusions are drawn 
in Section~\ref{concl} at the end of this work.

\section{Theory}  \label{Theory}

As a function of  temperature $T$,
the thermal attenuation of quantum features  in He atom spectra, such as elastic diffraction observed in angular distributions and the diffuse elastic peak observed in energy-resolved spectra, is given by a Debye-Waller (DW) factor. \cite{Hoinkes,Benedek-T,Hulpke, FariasR}  This is expressed as a multiplicative factor 
$\exp\{ -2 W({\bf k}_f, {\bf k}_i ,T)  \}$
where ${\bf k_i}$ and ${\bf k_f}$ are the incident and final wave vectors of the  He atom projectile.  This means that the intensity of any elastic peak is given by 
\begin{eqnarray} \label{Eq1-2}
I(T) ~=~ I_0 \, \exp\{ -2 W({\bf k}_f, {\bf k}_i ,T)\}  
~,
\end{eqnarray}
where $I_0$ is the intensity the peak would have at $T=0$ in the absence of zero-point motion (rigid lattice limit).   In general $I_0 > I(0)$.

The DW exponent is  expressed by
$2 W({\bf k}_f, {\bf k}_i ,T) = \left\langle (\Delta {\bf k}\cdot {\bf u}^* )^2  \right\rangle_T$,
where $\Delta {\bf k} = {\bf k}_f -{\bf k}_i$   is the scattering vector, ${\bf u}^*$ is the effective phonon displacement felt by the projectile atom upon collision, and $ \left\langle \cdot \cdot \cdot \right\rangle_T$  denotes the thermal average. However, He atom scattering experiments typically use energies  below 100 meV.  The atoms do not penetrate the surface, and in fact are exclusively scattered by the surface electron density a few~\AA~above the first  layer of atomic cores. Thus  the exchange of energy through phonon excitation  occurs via the phonon-induced modulation of the surface electron gas; in other words via the electron-phonon (e-ph) interaction. 
This implies that the effective displacement ${\bf u}^*$
is not that of the atom cores, such as would be measured in a neutron or X-ray diffraction experiment, but is the phonon-induced displacement 
of the electron distribution outside the surface at the classical turning point where the He atom is 
reflected.\cite{Manson-JPCL-16,Manson-SurfSciRep}  
However, the effective mean square displacement of the electron density is related directly to that of the atom cores and shares many of its properties.
Notably, for a crystal obeying the harmonic approximation and for sufficiently large temperature (typically temperatures comparable to or greater than the Debye temperature) 
$\left\langle (\Delta {\bf k}\cdot {\bf u}^* )^2  \right\rangle_T$
is to a very good approximation linear in $T$, and the proportionality between 
$\lambda_{HAS}$ and the DW exponent, for the simplest case of specular diffraction, reads as \cite{Manson-JPCL-16}
\begin{eqnarray} \label{Eq1}
2W({\bf k}_f, {\bf k}_i,T) ~=~ 4 \mathcal{N}(E_F) ~ \frac{m E_{iz}}{m_e^* \phi} ~ \lambda_{HAS} ~ k_B T
~,
\end{eqnarray}
where $\mathcal{N}(E_F)$  is the electronic density of states (DOS) per unit cell at the Fermi energy $E_F$, $m_e^*$ is the electron  effective mass, $\phi$ is the work function, $m$ is the projectile atomic mass and $k_B$ is the Boltzmann constant.
The quantity
$E_{iz} = E_i \cos^2(\theta_i) = \hbar^2 k_{iz}^2/2m$ is the incident energy associated with motion normal to the surface at the given incident angle  $\theta_i$.  For application to non-specular diffraction peaks or to other elastic features Eq.~(\ref{Eq1}) should be adjusted to account for the correct scattering vector appropriate to the experimental scattering configuration, namely $4 k_{iz}^2 \longrightarrow \Delta {\bf k}^2 = ({\bf k}_f - {\bf k}_i)^2$.

Given the form of Eq.~(\ref{Eq1}) it is useful to define the dimensionless quantity $n_s$ as
\begin{eqnarray} \label{Eq1x}
n_s ~=~ \frac{\pi \hbar^2 \mathcal{N}(E_F) }{m_e^* a_c}
~,
\end{eqnarray}
where $a_c$ is the area of a surface unit cell.
With this definition and the help of Eqs.~(\ref{Eq1-2}) and~(\ref{Eq1}) the following form for 
$\lambda_{HAS}$ is obtained
\begin{eqnarray} \label{alphax}
\lambda_{HAS} ~=~ \frac{\pi}{2n_s} \alpha \, ; ~~~\alpha ~\equiv~ \frac{\phi~ \ln[I(T_1)/I(T_2)]}{a_{c} k^2_{iz} k_B (T_2-T_1)}  
~,
\end{eqnarray} 
where $T_1$ and $T_2$ are any two temperatures in the linearity region.
Eq.~(\ref{alphax}) neatly separates the surface electronic properties 
from the quantities measured in an actual experiment.  The electronic properties
expressed by the ratio
$\mathcal{N}(E_F)/m_e^*$  are contained in the dimensionless $n_s$, while
the readily determined  work function of the surface and the experimentally measured slope of the DW exponent are  in $\alpha$.

In the case of supported graphene there is certain amount of charge transfer to and from the Dirac cones, depending on the difference between the work function of graphene and that of the substrate. The work function of self-standing graphene of 4.5 eV  is smaller than that of metal surfaces considered here. \cite{Giov-08} Therefore they should all act as acceptors. Angle resolved photoelectron scattering (ARPES) data show, however, that Ir(111) (H. Vita {\em et al.}), \cite{Vita-14} Pt(111) (P. Sutter {\em et al.}), \cite{Sutter-09} and Ni(111)  (A. Alattis {\em et al.}), \cite{Alattas-16} act as acceptors with respect to graphene, whereas Ru(0001)  (Katsiev {\em et al.}), \cite{Katsiev-12} and Cu(111)  (Walter {\em et al.}) \cite{Walter-11} act as donors. 
Former HAS studies on Gr/Ru(0001) have provided evidence that the tail of the substrate electron charge density
actually extends beyond the graphene. \cite{Farias-Carbon-15-2} 
Thus the substrate surface work function has been used in Eq.~(\ref{alphax}), its role being to account for the steepness of the He-surface repulsive potential within the WKB approximation. \cite{Manson-JPCL-16,Manson-SurfSciRep}
In either electron or hole doping each Dirac cone contributes a DOS at the Fermi level, 
$\mathcal{N} (E_F) = a_c k_F/ \pi \hbar^2 v_F^2$,
with $k_F$ the Fermi wavevector referred to a K-point of the surface Brillouin zone (SBZ), and $v_F$ the Fermi velocity. By identifying $m_e^*$ with the cyclotron effective mass for doped graphene $m_e^*=\hbar k_F/v_F$, \cite{Ariel-12} this gives $n_s = 1$ for a single Dirac cone. Although only one third of a cone is within the SBZ, so that diametric electron transitions at the Fermi level connect points in neighboring SBZs, the latter are equivalent to umklapp transitions with ${\bf G}$ vectors in 
the $\Gamma $M
directions between different cone thirds inside the SBZ (umklapp intervalley transitions). With the inclusion of these transitions, $n_s = 6$ is the appropriate value. 
      
It is noted however that these transitions couple to phonons near the zone center and give therefore a modest contribution to $\lambda$. A thorough ARPES study by Fedorov {\em et al.}~of graphene on Au(111), doped with alkali and Ca donor impurities, and electron concentrations ranging from $2 \cdot 10^{14}$  (for Cs) to $5 \cdot 10^{14}$ electrons/cm$^2$ (for Ca), shows that the major contribution to $\lambda$ actually comes from phonons near the zone boundaries KMK'. \cite{Fed14} The derived Eliashberg function, providing the e-ph-weighed phonon DOS projected onto the impurity coordinates exhibits resonances with graphene phonons whose wavevectors arguably correspond to good Fermi surface nestings. The fact that the separation between the ZA and the optical phonon peaks increases with doping from that for Cs, near the K-point, to the one at smaller wavevectors for the largest Ca doping suggests that the e-ph coupling is mostly due to KK' intervalley transitions, either direct along the six edges (thus counting 3) or umklapp along the three long diagonals. 
First principle calculations by Park {\em et al.} appear to confirm the role of these transitions. \cite{Par}
This is just the intervalley e-ph coupling mechanism introduced by Kelly and Falicov in the 1970s for charge density wave transitions in semiconductor surface inversion layers. \cite{Falicov-1,Falicov-2,Falicov-3} In this case $n_s = 6$ enumerates the different nesting conditions contributing to the e-ph coupling strength $\lambda$, and is here adopted for graphene and graphite, independently of whether they are electron- or hole-doped. 

In addition to the work of Federof {\em et al.} mentioned above, \cite{Fed14}
Calandra and Mauri have demonstrated that three optical branches of relatively large energy of 160 meV, located at the ${\bf Q}$ points of the Dirac cones, make the major contributions to the total e-ph interaction in supported graphene. \cite{Cal-07}
This agrees with the Eliashberg function for quasi-free standing K-doped graphene on Au reported by Haberer {\em et al.} from an ARPES 
analysis, \cite{Hab13}
which is concentrated in the optical region between 150 and 200 meV in both 
$\overline{\Gamma} \, \overline{\mbox{K}}$
and $\overline{\mbox{K}} \, \overline{\mbox{M}}$ directions (in the latter with minor contributions from the acoustic modes). Since all available HAS measurements have been performed around values of $k_B T$ well below the optical phonon energies, the usual high-$T$ equation (\ref{Eq1}) for the Debye-Waller exponent should be rewritten with its full temperature dependence as \cite{Manson-SurfSciRep}
\begin{equation} \label{Eq1-1}
2W({\bf k}_f, {\bf k}_i,T) ~=~ 4 \mathcal{N}(E_F) ~ \frac{m E_{iz}}{m_e^* \phi} \frac{1}{3N}
\sum_{{\bf Q}, \nu} ~\hbar \omega_{{\bf Q}, \nu} \left\{    
n_{BE}(\omega_{{\bf Q}, \nu}, T) + \frac{1}{2}
\right\} ~\lambda_{{\bf Q},\nu} 
~,
\end{equation}
where   
$n_{BE}(\omega_{{\bf Q}, \nu}, T)$ is the Bose-Einstein occupation number.
By considering in a first approximation only the contributions from the optical modes, all assumed to have the same average frequency $\omega_0$ (Einstein model), Eq.~(\ref{Eq1-1}) (with the definition  
$\lambda_{HAS} = \sum_{{\bf Q},\nu} \lambda_{{\bf Q},\nu}/3N$)
becomes
\begin{eqnarray} \label{Eq1-1-2}
2W({\bf k}_f, {\bf k}_i,T) = 4 \mathcal{N}(E_F) ~ \frac{m E_{iz}}{m_e^* \phi}
~ \hbar \omega_{0} \left\{    
n_{BE}(\omega_{0}, T) + \frac{1}{2}
\right\} \lambda_{HAS} 
~.
\end{eqnarray}
Clearly, in the high temperature limit 
where $k_B T >> \hbar \omega_0$
Eq.~(\ref{Eq1-1-2}) becomes identical with the previous Eq.~(\ref{Eq1}), but for arbitrary temperature Eq.~(\ref{alphax}) becomes
\begin{eqnarray} \label{alphax-1}
\lambda_{HAS} ~=~ \frac{\pi}{2n_s} \alpha \, ; ~~~\alpha ~\equiv~ \frac{\phi~ \ln[I_0/I(T)]}{a_{c} k^2_{iz} \hbar \omega_0 \left\{    
n_{BE}(\omega_{0}, T) + \frac{1}{2}
\right\}}  
~.
\end{eqnarray} 
In the high temperature limit Eq.~(\ref{alphax-1}) coincides with Eq.~(\ref{alphax}), while in the low temperature limit $k_B T << \hbar \omega_0$
it is again independent of $T$ and given by
\begin{eqnarray} \label{alphax-2}
\lambda_{HAS} ~=~ \frac{\pi}{n_s}  \frac{\phi~ \ln[I_0/I(0)]}{a_{c} k^2_{iz} \hbar \omega_0 }  
~.
\end{eqnarray} 
Since $k_B T < \hbar \omega_0$ the above will produce values of $\lambda_{HAS}$ that are smaller than that of Eq.~(\ref{alphax}).  Note, however, that the density of states $\mathcal{N}(E_F)$ is unaffected by this Einstein mode approximation, thus the appropriate choice remains $n_s = 6$.

The high temperature harmonic approximation for $\lambda_{HAS}$ of Eq.~(\ref{alphax}) is valid when the average phonon energy sampled by the e-ph interaction is lower than $k_B T$ over the temperatures at which the DW exponent is measured.  In general this condition is met for soft materials such as simple metals.  On the other hand one could call hard materials those for which high energy optical phonons are the dominant contributors to the e-ph interaction.  Representative of such hard materials are the supported graphene systems considered here, and the equal frequency approximation of Eq.~(\ref{alphax-1}) is more appropriate.

A value of 160 meV shall be used for $\omega_0$, as calculated by Calandra and Mauri  for the optical modes of supported graphene around the K-point of the surface Brillouin zone, i.e., the ones which mostly sustain multivalley coupling. \cite{Cal-07} The following discussion will show that the use of the high-T approximation of Eq.~(\ref{alphax}), currently used in the analysis of soft materials, probably implies a substantial overestimation of $\lambda_{HAS}$, whereas the Einstein approximation for hard materials at comparatively low temperature yields better agreement with the existing estimations of $\lambda$ from various other methods.

\section{Graphite}  \label{Graphite}

\begin{figure*}
\begin{center}
\includegraphics[width=0.7\textwidth]{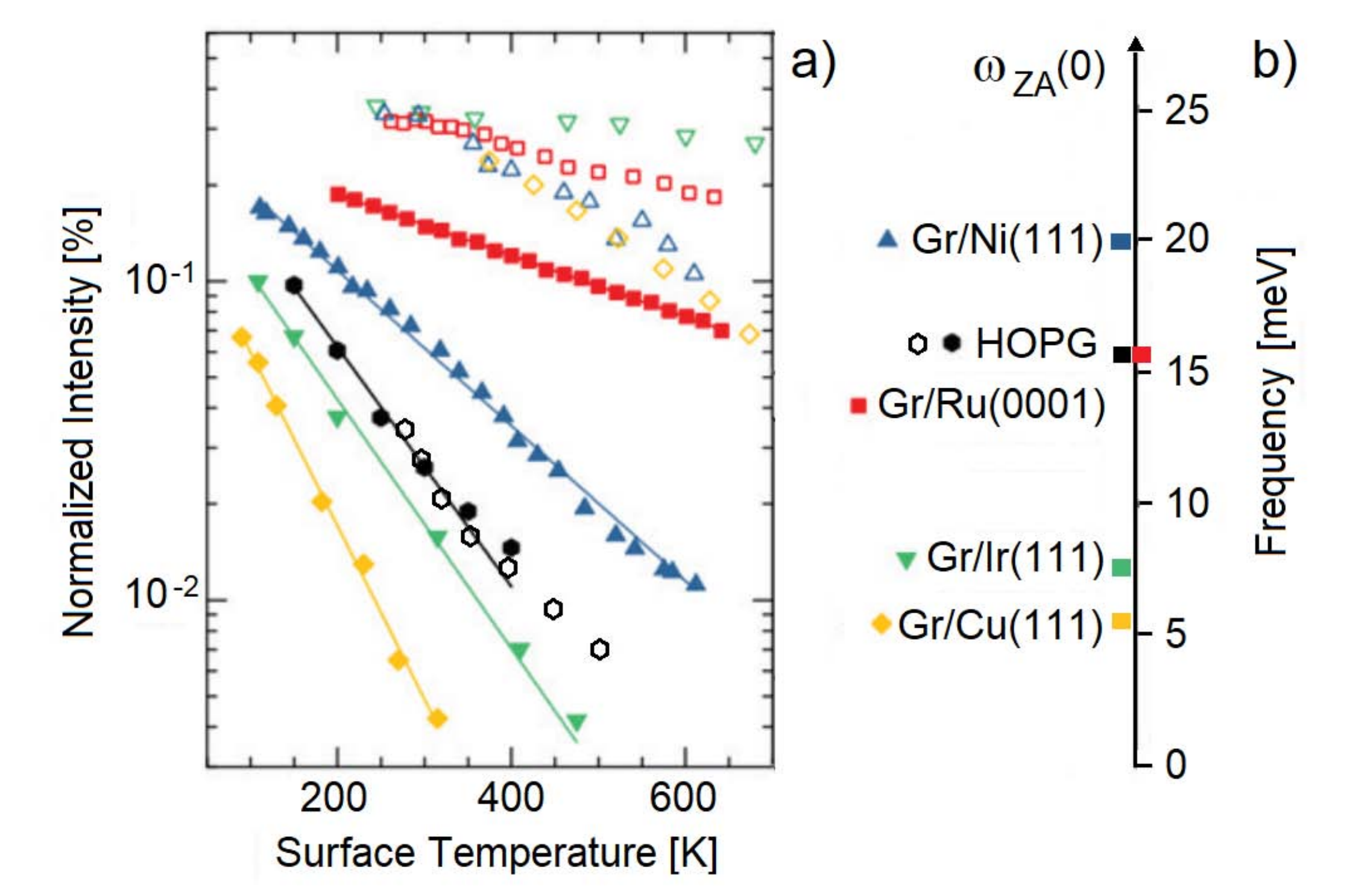}
\end{center}
\caption{ a) The normalized specular HAS intensity in log scale (proportional to the DW exponent) as a function of temperature for several systems of single-layer graphene supported on metal substrates (full symbols, defined in panel b)), together with similar data for the clean metals (open symbols). Also shown are the data for HOPG graphite at low (full hexagons \cite{Farias-2DMat-18}) and high temperature (open hexagons \cite{Vollmer}). b) The ZA mode frequency at ${\bf Q} = 0$ for HOPG graphite and all the Gr/metal systems shown in panel a).  The data are from Ref.~\cite{Farias-2DMat-18}
with some adjustment as explained in the text.}
\label{FigGr-metals-2}
\end{figure*}

Clean graphite C(0001) presents a weakly corrugated surface potential to He atom scattering which means that the specular peak is the dominant elastic scattering feature.  The thermal attenuation of C(0001)
has been measured by three independent groups over temperatures ranging from below 150 K to  500 K. \cite{Farias-2DMat-18,Kondo-09,Yamada-15,Vollmer} 
As is apparent from the plots of the DW exponent in Fig.~\ref{FigGr-metals-2}a), measured for graphite as a function of $T$ by Oh {\em et al.} \cite{Kondo-09} and by Vollmer \cite{Vollmer}, the slope is essentially linear at lower temperature up to about 350 K. 
The slopes of the different measurements, and using $n_s=6$, provide the values of $\lambda_{HAS}$ listed in Table~\ref{Gr-metals}
together with the input parameters and the respective references. 
The average over the available data for graphite gives from Eq.~(\ref{alphax}) (high-T limit) $\lambda_{HAS} = 0.36 \pm 0.09 ~(0.41 \pm 0.04~\mbox{below 400 K})$ or, from Eq. (7) (Einstein model at the average experimental temperature) 
$\lambda_{HAS} = 0.12 \pm 0.03 ~(0.13 \pm 0.02~\mbox{below 400 K})$.
In general, the indicated experimental error is that in the slopes of the DW plots, 
because uncertainties in other parameters extracted from the literature are not always given. However, our estimation is that
the overall uncertainty on $\lambda_{HAS}$ is around 15\% or less.

As is seen in Fig.~\ref{FigGr-metals-2}a), above 400 K the slope of $2W(T)$ clearly decreases, apparently tending to a value about a factor of 2 smaller at high temperature,  and $\lambda_{HAS}$ takes the values 0.20 from Eq.~(\ref{alphax}) and 0.09 from Eq.~(\ref{alphax-1}).  While with the high-T formula the reduction with respect to data below 400 K is by a factor 2, with the Einstein model the reduction is about 40\%. Thus the slope reduction at higher temperature is not due entirely to the different approximation, but is to a large extent an intrinsic effect. 
This behavior is very striking and is probably the consequence of the gradual transition of graphite from negative to positive in-plane thermal expansion, 
occurring at $\sim 500$ K. \cite{Bailey-70,Marsden-18}
A bond contraction, like that produced by an external pressure, generally yields a larger $\lambda$, while a dilation means a smaller $\lambda$, with similar consequences in superconducting 
materials on the critical temperature $T_c$. \cite{Bose-02}
This may explain the transition with temperature from a larger to a smaller 
$\lambda_{HAS}$ observed in graphite.  Interestingly, this is not observed in graphene, where the in-plane thermal expansion is predicted to remain negative up to a considerably larger temperature, well above the temperature range so far considered in HAS experiments. \cite{Mounet-05}

This behavior of two different slopes of the DW exponent observed in graphite, i.e., a steep slope at low temperatures and a less steep slope at high temperatures, is not seen in any of the supported graphene systems discussed below in the next Section.
It is also seen from the last column in Table~\ref{Gr-metals} that there is a rather large range of
reported values of $\lambda$ for bulk graphite, so that the values of $\lambda_{HAS}$ at the surface as measured by He atom scattering are well within the range of reported bulk values.

\section{Graphene on Metal Substrates}  \label{Graphene}

All of the systems of single layer graphene supported on close-packed metal surfaces for which DW plots of the specular diffraction peak have been measured are listed in Table~\ref{Gr-metals}.
For two of these systems, Gr/Ni(111) and Gr/Ru(0001), measurements were made with both He and Ne atom scattering.  For Gr/Ni(111), in addition to experiments done with He and Ne atoms, measurements were taken at very low incident energy using the $^3$He isotope
in spin-echo spectrometry.

\begin{table*}
	\caption{  
The e-ph coupling constants $\lambda_{HAS}$ obtained from atom scattering data for HOPG graphite and single-layer graphene adsorbed on metal crystal substrates are shown in the columns marked as $\lambda_{HAS}$ for the two calculational models.  All measurements were done with ordinary HAS, except where otherwise noted in the first column.
All values of $\lambda_{HAS}$ for Gr/metals were derived using  $n_s = 6$, and 
are calculated using the unit cell area $a_c$  of  the metal substrate.
The column marked as ``$\lambda_{HAS}/$substrate'' gives the values determined for the clean metal substrate surface.
(Values of other parameters showing no error indication are taken from the literature, where possible error sources are discussed. \cite{McMillan-68,Grimvall,k,AllenT,ref22,h})
A theoretical calculation by Park {\em et al.} produces a  value of up to $\lambda = 0.21$ for single-layer free-standing graphene, depending on doping level.\cite{Par} }
	\vspace{0.5cm}
	\begin{tabular}{|c||c|c|c|c|c|c|c|c|c|c|}
		\hline      Surface   & $T$
		& $ E_i$  & $ \phi $ &$\alpha$  & $ \lambda_{HAS}$ 
		&$ \lambda_{HAS}$ &$ \lambda_{HAS}$ & $\lambda$ (substrate   \\
		& $[K]$  & [meV]& $\left[ \mbox{eV} \right]$ &   
		& Eq.(\ref{alphax}) & Eq.(\ref{alphax-1})
		& substrate 
		&  bulk values)  \\
		\hline
		\hline C(0001)        &150-400 \cite{Kondo-09,Farias-2DMat-18} &  63  &
		4.5  \cite{Moos} & 1.70   & 0.41 & 0.12  &   &   0.70$\pm$0.08 \cite{Sug} \\
		& 250-360 \cite{Vollmer} &  69 & & 1.41    & 0.37 & 0.12 &  &0.034-0.28 \cite{Par} \\
		& 400-500 \cite{Vollmer} &  69 & & 0.76    & 0.20 & 0.09 &  & $\le0.20$ \cite{Leem}  \\
		& 300-360 \cite{Yamada-15} &  63 & & 1.76    & 0.46 & 0.16 &  &  \\
		\hline Gr/Ni(111)       & 200-400 \cite{Farias-17x,Farias-2DMat-18}  & 66  & 5.35 \cite{k}
		& 0.71  & 0.19 & 0.06 & 0.56 & 0.3, 0.7~~\cite{ref22} \\
		($^3$He atoms)       & 200-700 \cite{Tam-15} & 8  &
		& 0.69  & 0.16  & 0.08 &0.36  &  \\
		(Neon atoms)      & 100-200 \cite{Farias-17x}  & 66  &
		& 0.58  & 0.16  & (0.3) &  &  \\
		\hline Gr/Ru(0001)       & 300-400 \cite{Yamada-15}  & 67  &  4.71~\cite{k}
		&0.44   & 0.14  & 0.05 & 0.44  & 0.45~~\cite{AllenT} \\
		&200-600 \cite{Farias-Carbon-16}  & 32 &  & 0.38  &0.15  &0.06 &0.33 & 0.4 \cite{Grimvall} \\
		
		(Neon atoms)   & 90-150 \cite{Farias-17} & 43 &  &0.58  & 0.18  & (0.02) & 0.39 & \\
		\hline Gr/Ir(111)        & 100-500 \cite{Farias-2DMat-18} & 17.5  &
		5.76 \cite{k}  &1.88  & 0.54  & 0.17 & 0.30 & 0.41 \cite{ref22}, 0.34 \cite{McMillan-68}  \\
		& & & & & & & &0.50 \cite{AllenT} \\
		\hline
		Gr/Rh(111)        & 150-450 \cite{Sibener-16} &  19.3 & 4.98 \cite{k}
		& 1.65  & 0.43  & 0.14 &  & 0.41, 0.51 \cite{ref22} \\
		& &  63 &   & 1.16  & 0.31  & 0.10 &  &  \\
		\hline
		Gr/Pt(111)       & 300-400  \cite{Yamada-15} & 67  &
		5.70 \cite{k} & 1.53 & 0.54  & 0.20 &  & 0.66  \cite{ref22} \\
		\hline
		Gr/Cu(111)       &100-300 \cite{Farias-2DMat-18}  & 28  &4.98 \cite{k}
		& 2.56   &  0.69  &  0.22 & 0.083  & 0.093\cite{h},0.13 \cite{AllenT} \\
		& & & & & & & &0.1 \cite{Grimvall} \\
		\hline
	\end{tabular}
	\label{Gr-metals}
\end{table*}

In all these experiments, atom diffraction and diffuse elastic scattering measurements provide accurate control of the surface long-range order and defect concentrations, respectively, which ensures a high quality of the graphene-substrate interface.
The input parameters needed for evaluating $\lambda_{HAS}$  are given in Table~\ref{Gr-metals} together with the relevant references.  For several of the clean metals $\lambda_{HAS}$ has
also been independently determined and those values appear in the next-to-last column.  As stated previously for the case of graphite, the final column gives values from other sources  for the e-ph constant $\lambda$ of the pure bulk metal.  

For each of the Gr/metal systems two different calculations 
for $\lambda_{HAS}$
are exhibited in Table~\ref{Gr-metals}.  These are the two columns marked $\lambda_{HAS}$/Eq.~(\ref{alphax}) and $\lambda_{HAS}$/Eq.~(\ref{alphax-1}).  As denoted by the equation numbers, these calculations were carried out similarly to those explained above in Sec.~\ref{Graphite}, using $n_s = 6$ and, for Eq.~(\ref{alphax-1}) $\hbar \omega_0 = 160$ meV
as suggested by the Calandra and Mauri work. \cite{Cal-07}
The values for the Einstein model of Eq.~(\ref{alphax-1}) are around one-third to one-half those of the corresponding calculation using the high temperature limit of Eq.~(\ref{alphax}).

There is a significant amount of spread in the reported values for the bulk metals, but the values of $\lambda_{HAS}$
derived here for Gr/metals
compare favorably, and particularly so for the values calculated with the high temperature limit equation. It should  also be noted that those values of $\lambda_{HAS}$ are quite similar regardless of whether the projectile is He or Ne. The value 0.16 obtained for Gr/Ni(111) using
Ne atoms or
very low energy (8 meV) $^3$He atoms with the spin-echo detection technique is close to that of $^4$He at the much larger energy of 66 meV which is 0.19.
For the two cases for which data is available for Ne scattering, namely Gr/Ni(111) and Gr/Ru(0001), the calculated values using the Einstein model are placed in parenthesis.  This is because the Einstein model is probably not as valid because the heavy Ne projectiles are expected to excite a larger range of phonons than the lighter He atom projectiles.

\begin{figure*}
\begin{center}
\includegraphics[width=0.5\textwidth]{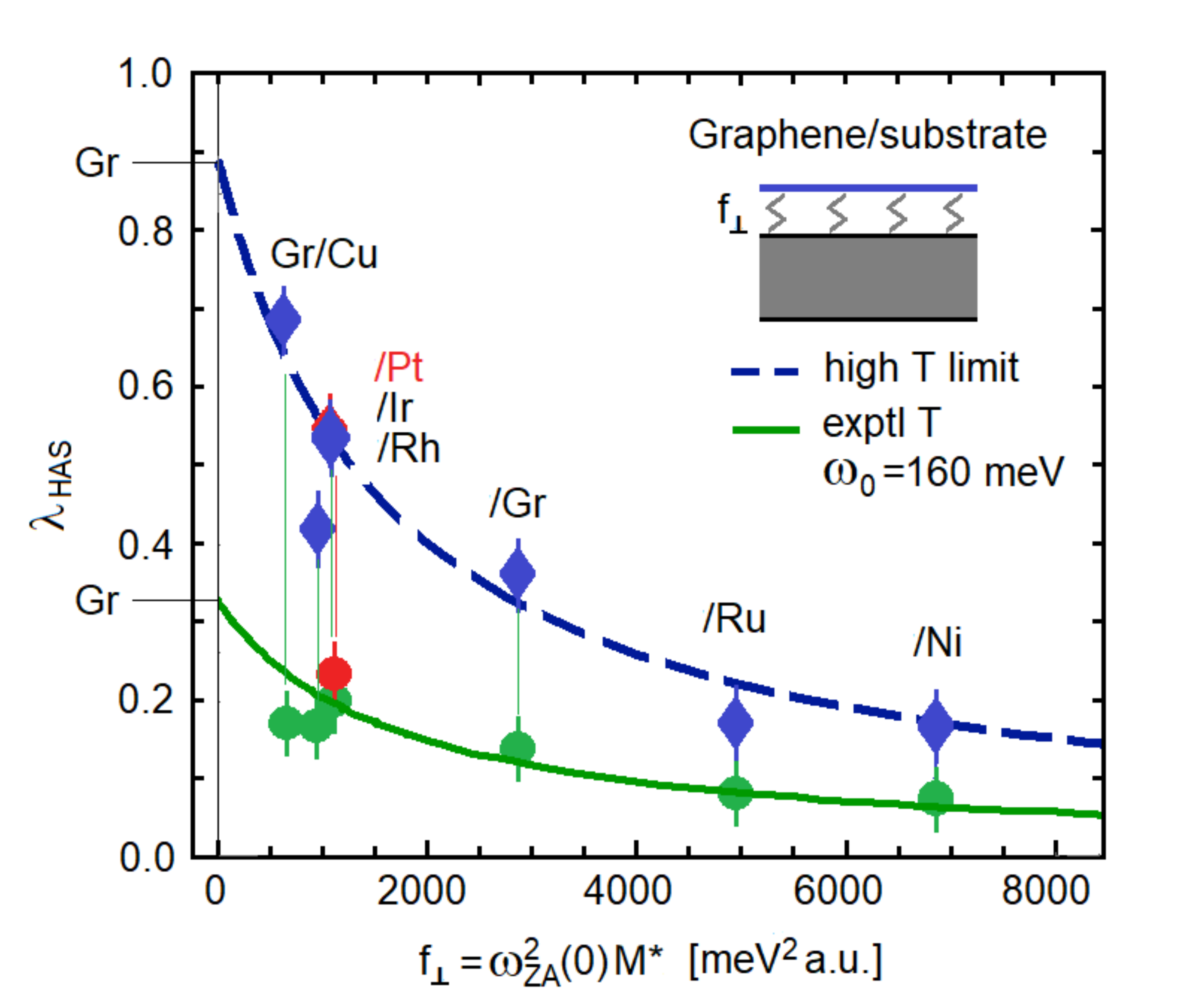}
\end{center}
\caption{The electron-phonon coupling constant $\lambda_{HAS}$ plotted as a function of the ZA mode spring constant $f_\perp$ coupling graphene to the substrate.  
The data are from Ref.~[\cite{AF16}], with some adjustment; see discussison in the text.
The upper dashed-line  fitting curve  with diamond-shaped data points, as explained in the text, allows the extrapolation of the value $\lambda = 0.89 \pm 0.04$ 
using the high temperature limit of Eq.~(\ref{alphax})
for ideally flat free-standing graphene (error bar calculated from the mean-square relative deviation from the fitting curve).
Similarly, the lower solid-line curve with circular data points is the result of the Einstein mode approximation of Eq.~(\ref{alphax-1}) with $\omega_0 = 160$ meV using the average experimental temperatures and which predicts the value of $\lambda = 0.32 \pm 0.09$ for free standing graphene.
}
\label{FigLambdaVsEza-3}
\end{figure*}

It is of interest to compare the values of $\lambda_{HAS}$ calculated here with the bonding strength of the graphene to these metals.  The Gr/metal bonding strength is usually judged by the Gr-metal separation distance, and the systems that have  been investigated fall into two different categories, weakly bonded with a separation distance greater than 3~\AA~and strongly bonded whose separation distance is 2.5~\AA~ or less.  Among the former are the close packed surfaces of Ag, Au, Cu, Pt and Ir, while the latter examples include Pd, Rh, Ru Ni, Co and Re. \cite{AF16} 
Another way of evaluating bonding strengths is by comparing the frequencies
$\omega_{ZA}(0)$ of the shear vertical ZA mode at the zone center at parallel wavevector ${\bf Q} = 0$.  At long wavelengths the ZA mode is nearly dispersionless as a function of ${\bf Q}$ for small ${\bf Q}$ and acts like an Einstein mode with a spring constant  $ f_\perp = M^*\omega_{ZA}^2(0)$  where  
$M^* = 2 M_C M_S/(2M_C + M_S)$ is the effective mass, $M_C$ the carbon mass and $M_S$ is the mass of the substrate atom.  
For graphite $M_S$  coincides with the planar unit cell mass $2 M_C$ and therefore $M^* = M_C$.  
Figure~\ref{FigLambdaVsEza-3} shows $\lambda_{HAS}$ plotted as a function of the interplanar force constant
$f_\perp$ for both of the two different $\lambda_{HAS}$ calculations.
Values of $\omega_{ZA}(0)$ measured by HAS are taken from Refs.~[\cite{Farias-2DMat-18}] and~[\cite{AF16}], except for that of Gr/Pt(111), extrapolated from Politano {\em et al.}~high resolution electron energy loss spectroscopy (HREELS) data. \cite{Politano-12}

Figure~\ref{FigLambdaVsEza-3}  
shows a clear and interesting correlation between
$\lambda_{HAS}$ and the graphene-substrate interaction: the weaker is this interaction, the stronger is the e-ph coupling in graphene.  This may be qualitatively understood by considering that a stronger interaction with the substrate 
implies a substantial reduction of the vertical mean square displacement, and possibly also 
some localization of graphene free electrons.

Some comments are in order about two of the values of $\omega_{ZA}(0)$ for the ZA mode that have been reported in the literature.  In the case of Gr/Ru(0001) two widely different values of 16 and 27 meV have been reported. \cite{Farias-Carbon-15-2,AF16,WGoodman} The value of 16 meV was measured by HAS over a  range of parallel wavevector ${\bf Q}$ that clearly demonstrates its dispersion behaving quadratically at large ${\bf Q}$ as expected (see discussion below in Sec.~\ref{ZA-flex}).
The value of 27 meV measured with HREELS \cite{WGoodman} appears to be incorrectly assigned.  The reason for this assessment is that this mode also appears in HREELS measurements of the clean Ru(0001) surface, although at the slightly higher energy of a little over 30 meV (at the $\Gamma$ point) where it is identified as the optical mode S$_2$. \cite{S2Gr,Widdra02}  The reason for the small energy difference between 27 and 30 meV can be explained by the fact that the tightly bound Gr simply weights down the outermost Ru layer, giving it a slightly smaller energy. 
Thus, it appears that the value of 27 meV for Gr/Ru(0001) should be assigned to the S$_2$ mode and not to the ZA mode.  In Fig.~\ref{FigLambdaVsEza-3} the value $\omega_{ZA}(0) = 16$ meV  has been used.  

Another reported value of $\omega_{ZA}(0)$ that needs to be discussed is that of Gr/Rh(111) for which a value of 7 meV has been reported. \cite{Sibener-16}  This value would appear to be too small when compared with the other strongly bound Gr/metal systems such as Gr/Ni and Gr/Ru where the value 
is rather in the range of 16 - 20 meV.  Moreover, no upward dispersion is apparently observed, as would be expected.  It is unfortunate that no phonon measurements were reported above 12 meV for this system.
Other HAS measurements, for example for Gr/Ru(0001) and Gr/Ni(111) have shown that there  are a variety of
other modes and
features that can produce peaks in the energy-resolved spectra at energies below 10 meV, including backfolding of the Rayleigh mode due to the super-periodicity of the moir\'e  patterns induced on these strong binding graphene 
layers.  \cite{Farias-Carbon-15-2,Farias-Carbon-16}
In spite of this caveat about $\omega_{ZA}(0)$ for Gr/Rh(111), the corresponding value   for $f_\perp = 954 \, meV^2 \, a. u.$ is not far from fitting into the observed
$\lambda_{HAS}$ vs. $f_\perp$  correlations and this value has been included in the two curves shown in Fig.~\ref{FigLambdaVsEza-3}.

The charge density oscillations detected by HAS receive the largest contributions from the phonon shear-vertical (SV) mean square displacements of surface atoms, and the latter are inversely proportional in self-standing graphene to an effective shear-vertical force constant $F$ (equal to 1908 meV$^2$ a.u. from the analysis of Gr/Ru(0001) HAS data \cite{Farias-Carbon-15-2}). Thus it is reasonable to suggest that $\lambda$ for the supported graphene is inversely proportional to $F+ f_\perp$ and scales therefore as $\lambda_{Gr}/(1 + f_\perp/F)$, with $\lambda_{Gr}$ (which is the value of $\lambda_{HAS}$ for self-standing graphene) and $F$ as fitting parameters.  The best fit is obtained with $F = 1650$ meV$^2$ a.u., which is indeed quite close to the above value from the previous HAS study on Gr/Ru(0001), and $\lambda_{Gr} =$ 0.32 and 0.89 from the experimental temperature with the Einstein 
model of Eq.~(\ref{alphax-1}), and in the high-T limit of Eq.~(\ref{alphax}), respectively (see Fig.~\ref{FigLambdaVsEza-3}). 
Since graphene is a diatomic lattice with two equal masses in the unit cell, the dispersion curve of the SV modes (as well as of the shear-horizontal (SH) ones) has a gapless folding point at the K-point of the acoustic into the optic branch with a single effective SV force constant. Thus the change from $F$ to $F+ f_\perp$  caused by the substrate equally affects both the acoustic and optical branches, producing a sizeable effect on $\lambda_{HAS}$.

Although the two different approximations of Eqs.~(\ref{alphax}) and~(\ref{alphax-1}) produce different values for $\lambda_{HAS}$ when projected to the limit of unsupported graphene, the similar nature of the two curves exhibited in Fig.~\ref{FigLambdaVsEza-3} shows that such calculations produce predictions of the behavior of $\lambda_{HAS}$ as a function of experimentally controllable parameters.  In this case the controllable parameter is the bonding strength to the metal substrate.  Predicting the behavior of the e-ph constant as a function of experimentally variable parameters is important, and a second example in which the parameter is doping concentration is discussed below.

The $\lambda_{HAS}$ values reported in Table~\ref{Gr-metals}  fall in the same range as those reported by Fedorov {\em et al.}~for electron-doping, \cite{Fed14} although the doping mechanism is different, one being achieved by changing the substrate, the other by changing the impurities. 
In both cases, however, $\lambda$ increases with the softening of an acoustic mode, whether respectively due to the ZA mode dependence on the force constant $f_{\perp}$ between graphene and the substrate, or to the same graphene ZA mode in resonance with the impurity-induced mode. Note that in the 
Fedorov 
{\em et al.}~experiments \cite{Fed14} the Cs and Ca doping levels mentioned above respectively correspond to about 0.05 and 0.13 electrons per C atom, which are in the range of present substrate-graphene charge transfers. Despite these similarities it should be noted that the impurity contribution to the Fermi level DOS is sufficiently low and can be neglected, whereas the substrate contribution can be relevant. In the extreme weak-coupling limit, unquenched substrate surface states can exist at the Fermi level, with an additional contribution to that of Dirac cones. The model parameter suitable to incorporate these contributions is $n_s$, which may then be greater than 6. It is therefore likely that the values of $\lambda_{HAS}$ reported for Gr/Cu(111) in Table~\ref{Gr-metals} and Fig.~\ref{FigLambdaVsEza-3} are overestimated. 
Actually it should be noted that all the values of $\lambda_{HAS}$ reported in Table~\ref{Gr-metals} for 
weakly bonded graphene systems
calculated with the high temperature limit
are much larger than the values for a graphene single layer as calculated by Park {\em et al.} even at comparatively large doping levels ($\lambda < 0.21$), \cite{Par} although consistent with values from recent studies on n-doped graphene discussed in the following paragraphs.

The possibility of increasing $\lambda$ with doping has stimulated several recent studies, all aiming at high-$T_c$ superconductivity in graphene. Just two examples are the analysis by Zhou {\em et al.}~on heavily N-(electron) and B-(hole) doped graphene, \cite{Zhou-15}  and the remarkable transport properties reported by Larkins {\em et al.}~in phosphorous-doped graphite and graphene, apparently suggesting the onset of superconductivity at temperatures as high as 260 K. \cite{Larkins-16} Among graphene systems where superconductivity is induced by the contact with a periodic Ca layer and the consequent addition of a 2D electron gas at the Fermi level, worth mentioning are the works by Yang {\em et al.}~where interband electron-phonon coupling is shown to play an important role, \cite{Yang-14} and by Chapman {\em et al.}~on Ca-doped graphene laminates. \cite{Chapman-16} Similarly, electron-doped material such as Li-covered graphene has also been predicted from first-principle calculations to attain a $\lambda$ value as large as 0.61 with a superconducting $T_c = 8.1$ K. \cite{Profeta}

Another way to induce superconductivity in graphene is the proximity effect, occurring, as demonstrated by Di Bernardo {\em et al.}~in a single layer graphene deposited on an oxide superconductor. \cite{Bernado-17}
The comparatively easy way to measure 
the e-ph coupling constant
with a HAS apparatus  using the temperature dependence of the DW exponent could certainly facilitate the search for novel superconducting graphene systems.

\section{Comparison of ZA and flexural phonon modes}  \label{ZA-flex}

When referring to graphene, an important remark is in order concerning the proportionality of $\lambda_{HAS}$ to the temperature-dependent mean-square displacements entering the DW exponent, due to a persistent confusion in the literature between the 
perpendicularly-polarized acoustic ZA modes and the flexural modes. The ZA modes are solutions of a thin elastic plate with fixed boundaries or periodic boundary conditions, and their frequency is proportional to the wavevector in the long-wave limit, whereas, flexural modes are solutions for a thin elastic plate with free boundaries and have their frequency in the long-wave limit as proportional to the square of the wavevector. The difference is simply due to the fact that transverse deformations with fixed boundaries imply a longitudinal strain, whereas in flexural modes the boundaries can move and no strain occurs.  

The difference between the ZA and flexural modes is particularly relevant for the DW factor because in a linear chain or a 2D lattice of $N$ atoms with cyclic boundary conditions the mean-square displacement at a given temperature tends to a finite value for $N \rightarrow \infty$, whereas in a linear chain or 2D lattice with free boundaries the mean-square displacement of the flexural mode diverges for $N \rightarrow \infty$ at any 
temperature. \cite{Peierls-34,Peierls-35,Landau-37,Landau-2,Mermin-66,Mermin-68}
Such large amplitude behavior has been reported on unsupported graphene as measured  with high resolution electron 
microscopy. \cite{Meyer-07,Zettl-07,Kis-2011,Tendeloo-11}
Moreover, exotic large amplitude vibrations at the free edges of suspended graphene ribbons, not predicted by classical homogeneous elastic plate dynamics, have been demonstrated by Garc\'ia-Sanchez  {\em et al.}~with a novel scanning probe microscopy method. \cite{Gar08}
Considering that experiments on perfectly suspended graphene flakes with free boundaries, as well as first-principle calculations without imposing cyclic boundary conditions, are hard to carry out, the frequent assumption of a quadratic dispersion for $\omega_{ZA}({\bf Q})$, and the term {\em flexural} for the ZA mode, appear to be inappropriate.

On the other hand, finite flakes of weakly coupled (quasi-self-standing) supported graphene may approximate the free-boundary condition, thus yielding a mean-square displacement rapidly increasing with temperature, i.e., a steep decrease of the DW factor. In such a case involving finite flakes, the association of DW slopes like those of Fig.~\ref{FigGr-metals-2}a) exclusively to  e-ph interaction may be incorrect, especially for the weakest graphene-substrate couplings.

In the context of measuring the electron-phonon constant, it is of interest to discuss the unique phonon modes added to the semi-infinite bulk when graphene is adsorbed onto the metal substrate.  A free-standing 
(unsupported) sheet of graphene has three  acoustic phonon branches, one with a longitudinal (LA) and two with transverse polarization, one in-plane (shear horizontal, SH) and one normal to the plane (shear vertical, SV or sometimes denoted as ZA).  By imposing cyclic boundary conditions as well as translational and rotational invariance conditions the three branches have a linear dispersion for ${\bf Q} \rightarrow 0$ corresponding to three distinct speeds of sound.  
When adsorbed to a substrate, the three modes lose their linear dispersion and at small parallel wave vector ${\bf Q}$ their frequencies go to finite values
$\omega_j(0)$ ($j =$ ZA, SH, LA) due to the bonding force constants between the graphene and the substrate.  Since for ZA modes the bonding force constant is essentially radial, while those for LA and SH modes are mostly transverse, in general $\omega_{ZA}(0) > \omega_{LA}(0)  > \omega_{SH}(0)  $.
The experimental $\omega_{ZA}(0)$, or better the corresponding bonding force constant $f_\perp$ can be taken as a measure of the graphene-substrate interaction.  For a free-standing infinite-extent thin membrane, such as graphene, the dispersion relation for the ZA mode exhibits a typical upward curvature due to the sp$^2$ bonding structure and consequent strong angle-bending forces, and can be approximated by \cite{Landau,Nishira,Balandin}
\begin{eqnarray} \label{ZA1}
\omega^2_{ZA}({\bf Q})  ~=~ v^2_{SV} {\bf Q}^2 ~+~ \frac{\kappa}{\rho_{2D}} {\bf Q}^4
~,
\end{eqnarray}
where $v_{SV}$ is the speed of SV waves, $\kappa$ is the bending rigidity (sometimes called the flexural rigidity), and $\rho_{2D}$ is the 2-D mass density.
However, if the membrane is coupled to the substrate, the coupling force constant will introduce a gap of frequency $\omega_{ZA}(0)$.  In this case the corresponding dispersion in the region of small ${\bf Q}$ is often 
expressed as \cite{Guinea-Amorim-13}
\begin{eqnarray} \label{ZA2}
\omega^2_{ZA}({\bf Q})  ~=~ \omega_{ZA}^2(0) ~+~ \frac{\kappa}{\rho_{2D}} {\bf Q}^4
~,
\end{eqnarray}
written without the quadratic term shown in Eq.~(\ref{ZA1}), its effect being negligible with respect to that of the other two terms.
It is Eq.~(\ref{ZA2}) that has been used to determine the bending rigidity $\kappa$ of the ZA mode of graphene supported by   the metals considered here \cite{Farias-Carbon-18,Sibener-16,Farias-Carbon-15,Farias-Carbon-15-2,Farias-Carbon-16} as well as for the thinnest known layer of vitreous glass, a bilayer of SiO$_2$ on Ru(0001). \cite{Holst-PRL}

\begin{figure*}
	\begin{center}
		\includegraphics[width=0.5\textwidth]{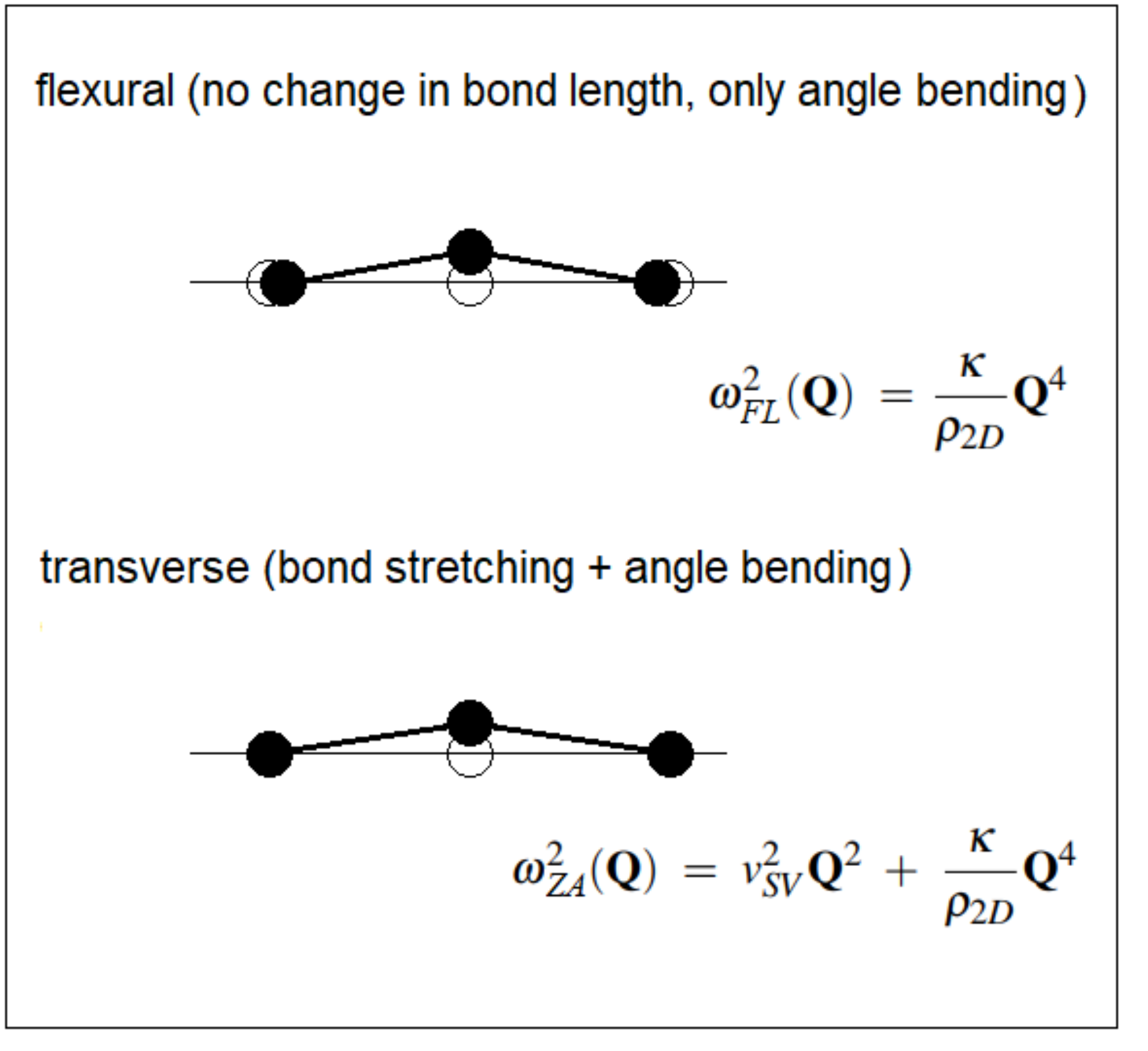}
	\end{center}
	\caption{ Angle bending displacement patterns in a linear chain for a flexural (FL) mode (above) and a shear-vertical (SV) transverse mode (below, also called the ZA mode), with the equations of the respective dispersion relations.  While the SV mode fulfills cyclic boundary conditions, which implies bond stretching, the FL mode does not and is only allowed by free boundary conditions.}
	\label{flexural}
\end{figure*}

It is important to note that most of the modelling of supported graphene dynamics is based on the assumption of a rigid substrate.  However, the phonon spectrum of graphene, with its large dispersion of acoustic modes and high-frequency optical modes, covers the whole phonon spectrum of the substrate, and therefore several avoided crossings are expected between graphene and the substrate modes of similar polarization.  For the ZA branch the important interactions are with the S$_2$ optical branch of the metal and the Rayleigh wave.  For this reason it was convenient to replace the carbon dimer mass with the effective mass in the expression of $f_\perp$ used in Fig.~\ref{FigLambdaVsEza-3} 
Moreover, supported graphene is no longer a specular plane, and coupled SV phonon modes acquire some elliptical polarization, leading, e.g., to an avoided crossing between the ZA and LA modes near ${\bf Q} =0$ (see, for example, HAS data for Gr/Ru(0001) \cite{Farias-Carbon-15-2}).

In recent works on free-standing graphene dynamics (and even on graphite dynamics) there is some apparent confusion between SV transverse (ZA) and flexural modes. \cite{Farias-Carbon-15,Dresselhaus,GuineaRMP,Rubio-04,Karssemeijer-11,Liu-Every-16}
This also appears in the case of other thin film materials such as bilayer SiO$_2$. \cite{Holst-PRL,Gao-Xie-Xu-16}
The study of the dynamics of elastic plates dates back to works of Euler, Bernoulli, D'Alembert, Sophie Germain and Lagrange, just to mention some of the pioneers, and the terminology is well established. \cite{Szilard-04}
When transferred to lattice dynamics, transverse modes refer to lattices with cyclic (or fixed) boundary conditions, whereas flexural modes refer to lattices with free boundaries.  The difference is illustrated for a three-atom chain in Fig.~\ref{flexural}.  The flexural mode, where angle bending occcurs with no change of the bond lengths, clearly does not fulfill cyclic boundary conditions, and the frequency dispersion is not linear for ${\bf Q} \rightarrow 0$, but is quadratic.  On the other hand, in the transverse ZA mode angle bending occurs with bond stretching in order to fulfill the boundary conditions, and the frequency dispersion is linear for ${\bf Q} \rightarrow 0$
as shown in Eq.~(\ref{ZA1}).

\section{Conclusions} \label{concl}

To conclude, what has been demonstrated here is that a relatively simple measurement of the thermal attenuation of He atom diffraction can be used to extract values of the e-ph coupling constant $\lambda= \lambda_{HAS}$ for single layer graphene supported on  close-packed metal substrates. This is an extension of previous work that determined $\lambda_{HAS}$ for simple metal surfaces, and shows that the e-ph coupling constant can be determined for much more complicated surface systems.  All available data for both He and Ne atom scattering from Gr/metals have been analyzed here, as well as data for clean graphite.  A significant result arises when $\lambda_{HAS}$ is plotted as a function of the spring constant $f_{\perp}$ binding the graphene to the substrate: the weakly bonded systems have large values of $\lambda_{HAS}$
while the strongly bonded systems have smaller values.  
In a plot of $\lambda_{HAS}$ versus $f_{\perp}$ all data points fall on a smooth curve 
according to the law $I_{HAS}(f_{\perp})  = I_{HAS}(0) / (1 + f_{\perp}/F)$ with $F$  the effective force constant of graphene.
This fit allows a very reasonable extrapolation to $f_{\perp} \rightarrow 0$, corresponding to free-standing graphene, and the two calculational models considered here give predictions that vary by a factor of nearly 3.
This interesting prediction is that a large, flat, free-standing sheet of graphene (obeying fixed or cyclic boundary conditions) with a free carrier concentration of the order of  10$^{13}$ cm$^{-2}$, like the supported graphene systems considered here, should have an e-ph constant with the relatively large value of about 0.89 as calculated with the high temperature limit of Eq.~(\ref{alphax})  or about 0.32 as given by the Einstein mode model of Eq.~(\ref{alphax-1}).
Although this difference in $\lambda_{HAS}$ values for free standing graphene may seem large, it should be noted that even for bulk materials different measurements of $\lambda$ produce even larger discrepancies, as for example is evident in the last column of Table~\ref{Gr-metals}.  The two different curves exhibited in Fig.~\ref{FigLambdaVsEza-3} are qualitatively similar, which points out that even though different calculational models may produce quantitative differences, both models describe similar trends as functions of controllable experimental parameters.  Thus calculations of $\lambda$ such as presented here could be useful for comparing different systems and, for example, within a given Gr-substrate system comparing the effects of doping or other ways to alter the surface electron density.

It would be extremely interesting to verify this free-standing graphene prediction by carrying out He atom scattering experiments on unsupported graphene.
The implications of this work suggest several additional experiments that would be of interest to carry out with He atom scattering on graphene and related systems.  One such class of experiments would be to measure extensive energy-resolved inelastic scattering spectra.  Such spectra exhibit peak-features due to specific phonons, and these peak intensities are directly proportional to $\lambda_{{\bf Q},\nu}$, the mode-selected e-ph contributions  to $\lambda_{HAS}$. \cite{Benedek-14} Thus, inelastic atom-scattering can provide unique information on which phonon modes contribute most importantly to $\lambda$.  

Another class of experiments would be to measure double and multiple layer supported graphene, and in particular it would be important to measure $\lambda$ for twisted bilayer graphene (tBLG) which can be superconducting for specific twist angles, \cite{Bis11,Cao18s,Cao18b,Mog20} as well as measurements on the
class of layered transition-metal chalcogenides which exhibit 2D superconductivity
(see, e.g., the recent work by Trainer {\em et al.} \cite{Trainer} and the HAS study of 2H-MoS$_2$ \cite{Anemone-19}). In the specific case of tBLG, where its peculiar electronic structure is considered to favor a strong electron correlation as the basic mechanism for pairing, \cite{Bis11,Cao18s,Cao18b,Liu18,Yan19} the actual value of $\lambda$ would be rather small. On the other hand, it has been suggested that the very same electronic structure at the Fermi level supports a strong e-ph interaction with $\lambda$ of the order of 1, \cite{Wu18a,Cho18,Das19,Wu19b} or even 1.5, \cite{Lia19} so as to consider tBLG as a conventional superconductor. This would rely on a strong multivalley e-ph interaction as well as on a decoupling between twisted layers, similar to the orientational stacking faults in graphene multilayers grown on 4H-SiC(0001) which makes the surface layer behave as a self-standing doped graphene.\cite{Has08} This may well correspond to the large $\lambda$ in the limit $f_{\perp} \rightarrow 0$ represented in Fig.~\ref{FigLambdaVsEza-3} as calculated with the high temperature model. 
Helium atom scattering on tBLG and twisted multilayer graphene would certainly help to clarify the above issue.

\vspace{1cm}
{\bf Acknowledgments}: We would  like to thank Profs.~M. Bernasconi, P. M. Echenique, E. Chulkov and D. Far\'ias for helpful discussions. This work is partially supported by a  grant with Ref. FIS2017-83473-C2-1-P from the Ministerio de Ciencia (Spain).
\vspace{1cm}
%












\end{document}